\documentclass{cimento}

\usepackage{graphicx}  

\title{Universality and Evolution of TMDs}
\author{S.~M.~Aybat\from{ins:nikhef}
        \atque
T.~C.~Rogers \from{ins:sb}}
\instlist{\inst{ins:nikhef}Nikhef Theory Group, Science Park 105, 1098XG Amsterdam, The Netherlands
            \inst{ins:sb} C.N. Yang Institute for Theoretical Physics, Stony Brook University, Stony Brook, NY 11794, USA}

\PACSes{\PACSit{12.38.Bx}{Perturbative Calculations.}
\PACSit{12.39.St}{Factorization.}
\PACSit{12.38.Cy}{Summation of Perturbation Theory.}}

\begin{document}

\maketitle

\begin{abstract}
In this talk, we summarize how QCD evolution can be exploited 
to improve the treatment of transverse momentum dependent (TMD) parton distribution and fragmentation functions.  
The methods allow existing non-perturbative fits to be turned into fully evolved TMDs that are 
consistent with a complete TMD-factorization formalism over the full range of $k_T$.  We argue that  evolution is essential to
the predictive power of calculations that utilize TMD parton distribution and fragmentation functions, especially 
TMD observables that are sensitive to transverse spin.
\end{abstract}

\section{Collinear versus TMD Factorization}
\label{sec:collvstmd}
The standard collinear QCD factorization theorems have set the standard for 
the use of perturbative QCD calculations to probe certain properties of the microscopic structure of matter.
It will be instructive to recall the essential ingredients that lend the standard factorization treatments their predictive power:
\begin{itemize}
\item
Unambiguous prescription for calculating pertrubatively well-behaved higher 
order corrections to the hard scattering.
\item
Correlation functions describing the non-perturbative factors are 
well-defined and have universality properties so that, once measured, they 
can be useful for future phenomenological studies.
\item 
Evolution equations, to allow the correlation functions to be compared at different scales.
\end{itemize}
The collinear PDFs and fragmentation functions have by now 
been parametrized by a wide range and variety of experimental data, and 
have become indispensable tools in general high energy physics.
Ideally, TMD-factorization, in which the PDFs and fragmentation functions also carry information about the 
intrinsic transverse momenta of partons, should follow a very similar framework.  
However, it is only very recently~\cite{collins,Ji:2004wu,Ji:2004xq} that TMD-factorization 
has  reached a level of logical completeness similar to the well-known collinear cases.  As long there was 
ambiguity in the formulation of TMD-factorization, the procedure for correctly identifying TMDs in fits, as well 
as the correct procedure for implementing evolution has 
remained unclear.  

In the absence of a full TMD-factorization treatment, there have been several popular but separate 
approaches for dealing with TMDs:
\begin{itemize}
\item Generalized Parton Model (GPM):  A phenomenological approach is to extract TMDs from 
fits to data while assuming a literal parton model interpretation of the TMDs.  
One typically ignores evolution and therefore the fits 
correspond to specific scales.  
\item Resummation:  Begin with a collinear treatment valid for large transverse momentum, and 
attempt to improve the treatment of lower transverse momentum by resuming logarithms of $q_T/Q$.
A severe limitation of this approach is that it is bound to fail below some $q_T$, where many of the most 
interesting effects of TMD physics are expected to become important.
\item Models:  Non-perturbative models of TMDs can be used to study specific non-perturbative 
aspects of hadron structure (see A. Bacchetta's talk for a review of model calculations.).  But there are ambiguities in how these TMDs are related to the 
ones used in actual perturbative QCD calculations of cross sections.  
In particular, it is unclear what hard scale they should correspond to. 
\item Lattice Calculations:  Lattice calculations of TMDs (see, e.g.,~\cite{Musch:2010ka}) describe the non-perturbative 
distribution of patrons from first principles, but also require a clear definition for the 
TMDs, and a clear prescription for use in complete cross section calculations.
\end{itemize}
A useful TMD-factorization treatment should allow the advantages of each of these approaches 
to be unified within a single, clear formalism for relating TMD studies to observable cross sections.  
Fortunately, this is now possible following the recent work of Ref.~\cite{collins} (see, esp., chapts. 10 and 13).  
(A similar general approach was developed earlier by Ji, Ma, and Yuan~\cite{Ji:2004wu,Ji:2004xq}, which built 
upon the Collins-Soper-Sterman (CSS)~\cite{CSS2,CSS3} formalism.)
We will show how fits to TMDs can be constructed from existing work that follows the above tabulated 
approaches, but which include evolution and are consistent with a full TMD-factorization.

We apply the TMD-factorization method developed recently by Collins in Ref.~\cite{collins}.  
The factorization theorem for the Drell-Yan (DY) process, for example, is 
\begin{eqnarray}
\label{dyhadronic}
W^{\mu\nu}_{\rm DY} &=& \sum_f |\mathcal{H}_f(Q^2,\mu)|^{\mu\nu}\int d^2{\bf k}_{1T} d^2 {\bf k}_{2T} F_{f/P_1}(x_1,{\bf k}_{1T},\mu,\zeta_{1F})F_{f/P_2}(x_2,{\bf k}_{2T},\mu,\zeta_{2F})\nonumber\\
&\ &\hspace{20mm}\times
\delta^{(2)}({\bf k}_{1T} + {\bf k}_{2T} - {\bf q}_T) + {\rm Y} + \mathcal{O}(m/Q)\,.
\end{eqnarray}
Note the similarity of the first term above to a GPM description; there is a hard part $\mathcal{H}_f(Q^2,\mu)|^{\mu\nu}$  and a convolution of 
two TMD PDFs, $F_{f/P_1}(x_1,{\bf k}_{1T},\mu,\zeta_{1F})$ and $F_{f/P_2}(x_2,{\bf k}_{2T},\mu,\zeta_{2F})$.  However, in the full 
TMD-factorization treatment, they have acquired scale dependence through the renormalization group parameter $\mu$ and $\zeta_{1F}$ and $\zeta_{2F}$
 (which obey $\sqrt{\zeta_{1F} \zeta_{2F}} = Q^2$).  Moreover, there is no explicit appearance of a soft factor.  
  The first term in Eq.~\ref{dyhadronic} is appropriate for describing the small  $q_T$ region ($q_T << Q$).
 The $Y$ term corrects the large $q_T$ behavior and can be calculated in terms of normal integrated PDFs.
 Ref.~\cite{collins} derives very specific operator definitions for the TMD PDFs, which include the role of soft gluons and account 
 for all spurious divergences which have hindered efforts to clearly define the TMD PDFs in the past.

\section{Evolution of TMDs}
\label{sec:tmdevolution}
The evolution of the individual TMDs in transverse coordinate space is governed by the Collins-Soper (CS) equation~\cite{CSS2},
\begin{equation}
\label{csevolution}
\frac{\partial \ln \tilde{F}(x,b_T,\mu,\zeta)}{\partial \ln \sqrt{\zeta}} = \tilde{K}(b_T,\mu)\,,
\end{equation}
and the renormalization group equations,
\begin{equation}
\label{rgevolution}
\frac{d\tilde{K}(b_T,\mu)}{d\ln \mu}=-\gamma_K(g(\mu))\,, \qquad \frac{d\ln\tilde{F}(x,b_T,\mu,\zeta)}{d\ln\mu}=-\gamma_F(g(\mu),\zeta / \mu^2).
\end{equation}
In Eq.~\ref{csevolution}, $\tilde{K}(b_T,\mu)$ is the kernel for evolution with respect the energy variable $\zeta$, while 
$\gamma_F(g(\mu),\zeta / \mu^2)$ and $\gamma_K(g(\mu))$ are the anomalous dimensions.  These can all be calculated 
in perturbation theory, though $\tilde{K}(b_T,\mu)$ becomes non-perturbative at large $b_T$ and needs to be fit.  It is a universal function, however, 
both with respect to different processes and with regard to PDFs versus fragmentation functions.

\section{Specific Fits}
\label{fits}

Working within a GPM approach, Gaussian parametrizations have been fit to 
various TMDs in, for example, Ref.~\cite{Schweitzer:2010tt} from low energy SIDIS measurements.  The non-perturbative information corresponding to $\tilde{K}(b_T,\mu)$ at large $b_T$ in 
Eq.~(\ref{csevolution}) has been extracted from more traditional applications of the CSS method, such as in Ref.~\cite{Landry:2002ix}.  
With the non-perturbative parts constrained, specific tables for the TMDs in Eq.~(\ref{dyhadronic}) can be generated for any arbitrary scale $Q$ once the 
perturbative contributions have been calculated.  

\begin{figure}
\vspace{22.5mm}
\centering
\includegraphics[scale=0.5]{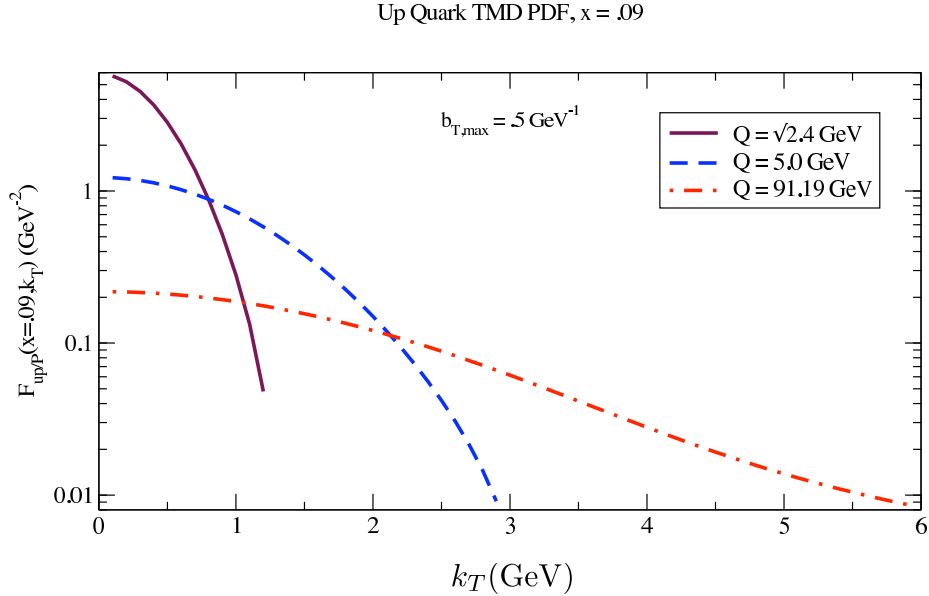}
\caption{Unpolarized TMD PDF for an up quark, for scales $Q=\sqrt{2.4}$, $5$ and $91.19$ GeV.
}
\label{TMDPDF}
\vspace{-5mm}
\end{figure}

We have carried this out explicitly for the unpolarized TMD PDFs and fragmentation functions in Ref.~\cite{Aybat:2011zv}, and have made 
the results available at  the website: \\ https://projects.hepforge.org/tmd/. \\
An example showing the unpolarized up quark PDF for a selection of $Q$ values is the graph in Fig.~\ref{TMDPDF}.

Note that, once the $Y$ term is included in Eq.~(\ref{dyhadronic}), the factorization formula exactly valid 
over the whole range of $k_T$ from $k_T=0$ to $Q$.

%%%%%%%%%%%%%%%%%%%%%%%%%%%%%%%%%
%%%%%%%%%%%%%%%%%%%%%%%%%%%%%%%%%

\section{Conclusions and Future Directions}
\label{sec:conclusions}
A crucial element of the TMD-factorization approach is that the TMDs are well-defined and, therefore, can be taken to be genuinely universal.
One consequence of this is that different theories of the non-perturbative properties of hadron structure can be compared with one another, 
and with experimental results over a range of hard scales.  

Ultimately, in order to take full advantage of the predictive power of TMD-factorization, requires a concerted effort to improve upon 
non-perturbative fits, to calculate higher orders to the anomalous dimensions, K, the coefficient functions, and the hard parts, and to 
provide numerical implementations of evolution in transverse momentum space similar to what has been done in Ref.~\cite{Aybat:2011zv}, but applied to other TMD PDFs and fragmentation functions, 
including interesting spin dependent ones.  We have recently been involved in efforts to extend evolution to the Sivers function~\cite{sivers}, and plan to continue 
this work to generate evolved fits for most of the various spin-dependent TMDs.

\begin{acknowledgments} 
M.~Aybat thanks the organizers of the Transversity 2011 workshop for the invitation to this interesting and fruitful workshop.
Support for this work was provided by the research 
program of the ``Stichting voor Fundamenteel Onderzoek der Materie (FOM)'', 
which is financially supported by the ``Nederlandse Organisatie voor Wetenschappelijk Onderzoek (NWO)''.
\end{acknowledgments}

\end{document}